\documentclass[aps,prb,twocolumn,showpacs,superscriptaddress,amsmath,amssymb]{revtex4}

\usepackage{graphicx}

\newcommand{\mathsym}[1]{{}}

\begin{document}

\title{Quantum model of an autonomous oscillator in hard excitation regime}

\author{E.D.~Vol}
\email{vol@ilt.kharkov.ua}
\affiliation{B.~Verkin Institute for Low Temperature Physics and Engineering NASU, 61103 Kharkov, Ukraine}

\author{M.A.~Ialovega}
\email{ialoveganicolas@gmail.com}
\affiliation{V.N.~Karazin Kharkov National University, 61077 Kharkov, Ukraine}

\received{\today}

\begin{abstract}
We propose the simple quantum model of nonlinear autonomous
oscillator in hard excitation regime. We originate from classical equations
of motion for similar oscillator and quantize them using the Lindblad
master equation for the density matrix of this system. The solution for the
populations of the stationary states of such oscillator may be explicitly found in the
case when nonlinearity parameters of the problem are small. It was shown
that in this situation there are three distinct regimes of behavior of the
model. We compare properties of this model with corresponding ones of close open system, namely quantum oscillator in soft excitation regime.
We discuss a possible applications of the results obtained.
\end{abstract}

\maketitle

The main goal of the paper is to introduce and consider the quantum model of
nonlinear autonomous oscillator (AO) in hard excitation regime. Our basic
tool for solving this problem is the Lindblad master equation (LME) which
describes the evolution of any (closed or open) Markov quantum system.
Clearly, the first aspiration that arises when one starts to study the
behavior of certain complex quantum open system (OS) is the desire to reduce
it to some more simple standard model that permits the rigorous mathematical
analysis. In the theory of open systems there are at least two similar
models namely 1) AO in soft excitation and 2) AO in hard excitation regimes.
The first case have been studied in Ref.~\onlinecite{1} where to this end the
semi-classical method of quantization of classical non-hamiltonian systems
was proposed. Therefore in the present paper we will focus our attention on
the case of AO in hard excitation regime. Note that AO both in soft and hard
excitation regimes are widely used in physics, biology and other sciences.
For example, in physics, an oscillator in soft excitation regime used as the
standard model of a generator of electromagnetic oscillations. As regards to
AO with hard excitation this system finds various applications aside from
physics as well for example in biology where similar model can be applied
for the description of activity of the giant axon of a squid in sea water~\cite{2}.
Now let us describe briefly the method of transition from known
classical equations of motion to quantum dynamics by means of the LME. The
basic idea in this way is the correspondence principle in the form proposed
by P. Dirac in his prominent book~\cite{3}.

It turns out that the broad interpretation of correspondence principle
allows one under certain conditions to quantize (at least in the
semi-classical approximation) the equations of motion not only for closed but
also for open systems using the LME which realizes the quantum description
of the evolution of quantum OS in the Markov approximation. This equation
for the evolution of the density matrix of quantum OS $\hat{\rho}$ has the
following general form (see~Ref.~\onlinecite{4}):

\begin{equation}
\frac{d\hat{\rho}}{dt}=-\frac{i}{\hbar }[\hat{H},\hat{\rho}%
]+\sum_{j=1}^{N}\{[\hat{R}_{j}\hat{\rho},\hat{R}_{j}^{+}]+[\hat{R}_{j},\hat{%
\rho}\hat{R}_{j}^{+}]\},
\label{1}
\end{equation}

where $\hat{H}$ is - an hermitian operator (Hamiltonian), which describes
the internal dynamics of quantum OS, and a set of non-hermitian operators $\{%
\hat{R}_{j},\hat{R}_{j}^{+}\}$- models its interaction with the environment.

The recipe of quantization proposed in Ref.~\onlinecite{1} consists of three
successive steps (its justification and all details see in this paper).
Firstly, the input dynamical equations should be presented in the special
form allowing the quantization (FAQ). In the simplest case of a system with
one degree of freedom with dynamical variables $x$ and $p$ or equivalently
with complex coordinate $z=\frac{x+ip}{\sqrt{2}}$ the desired equation in
FAQ looks as follows:

\begin{equation}
\frac{dz}{dt}=-\frac{i}{\hbar}\frac{dH}{dz^*}+\sum_{j=1}^N\{\bar R_j\frac{%
dR_j}{dz^*}-R_j\frac{d\bar R_j}{dz^*}\}. 
\label{2}
\end{equation}

This representation, in the case where it is found determines automatically
the classical functions $H(z,z^{\ast }),R(z,z^{\ast })$ and $\bar{R}%
(z,z^{\ast })$ entered in Eq.~(\ref{2}).

The second step is to find the quantum analogs of classical functions $\hat{H%
},\hat{R}$ and $\hat{R}^{+}$. To this end the simple rule can be proposed:
one should replace in all classical variables the coordinates $z$ and $%
z^{\ast }$ by the Bose operators $\hat{a}$ and $\hat{a}^{+}$. After this
procedure the operators $\hat{H},\hat{R}$ and $\hat{R}^{+}$ thus obtained
should be substituted into the LME. Now let us demonstrate in detail how the
method of quantization operates in the case of AO in hard excitation regime.
We will consider the simplest model of such oscillator that can be described
by the following equation of motion for the complex coordinate $z $ (see~ Ref.~\onlinecite{5}):

\begin{equation}
\dot z=-i\omega z- \varepsilon_1 z+\varepsilon_2 z |z|^2 - c z |z|^4, 
\label{3}
\end{equation}

where $\varepsilon _{1},\varepsilon _{2}$ and $c$ - are the constants,
describing the behavior of the oscillator. We will interested mainly in
possible stationary regimes of the behavior of the oscillator as functions
of these constants. One can easily verify that Eq.~(\ref{3}) can be represented in
the FAQ. Indeed let us introduce the functions $H=\omega z^{\ast }z,R_{1}=%
\sqrt{\varepsilon }_{1}z,R_{2}=\sqrt{\frac{\varepsilon _{2}}{2}}{z^{\ast }}%
^{2},R_{3}=\sqrt{\frac{c}{3}}z^{3}$. After that r.h.s. of Eq.~(\ref{3}) may be
written down as:

\begin{equation}
\begin{split}
-i\omega z- \varepsilon_1 z+\varepsilon_2 z |z|^2 - &c z |z|^4=-i\frac{dH}{
dz^*}{}\\&+\sum_{j=1}^3\{\bar R_j\frac{dR_j}{dz^*}-R_j\frac{d\bar R_j}{dz^*}\}.
\end{split}
\end{equation}

In what follows we will assume that $c=1$ since this case is always may be
achieved by choosing of appropriate time scale. According to the above
mentioned recipe of quantization the LME for the AO in hard excitation
regime takes the following form:

\begin{equation}
\frac{d\hat{\rho}}{dt}=-\frac{i}{\hbar }[\hat{H},\hat{\rho}%
]+\sum_{j=1}^{3}\{[\hat{R}_{j}\hat{\rho},\hat{R}_{j}^{+}]+[\hat{R}_{j},\hat{%
\rho}\hat{R}_{j}^{+}]\},
\label{5}
\end{equation}

where $\hat{R}_{1}=\sqrt{\varepsilon }_{1}\hat{a},\hat{R}_{2}=\sqrt{\frac{%
\varepsilon _{2}}{2}}\hat{a}^{+2},\hat{R}_{3}=\sqrt{\frac{1}{3}}\hat{a}^{3}$.

From physical reasons we expect that steady regimes of classical system~(\ref{3})
in quantum case correspond to stationary states of its quantum analogue
described by the LME~(\ref{5}). We will seek the stationary solutions of Eq.~(\ref{5}) in
the form $\hat{\rho}_{st}=\sum_{n=0}^{\infty }|n\rangle \rho _{n}\langle n|$%
, where $|n\rangle $ - are eigenvectors of the operator $\hat{n}$ or in
other words we assume that $\hat{\rho}_{st}$ is a certain function of
operator $\hat{n}$. Using the standard rule of commutation: $[\hat{a},\hat{a}%
^{+}]=1$ after the simple algebra we obtain the following difference
equation for the unknown coefficients $\rho _{n}$:

\begin{equation}
\begin{split}
2\varepsilon_1&((n+1)\rho_{n+1}-n\rho_n)+\varepsilon_2(n(n-1)\rho_{n-2}{}\\
&-(n+2)(n-1)\rho_n)+((n+3)(n+2)(n+1){}\\&
\times\rho_{n+3}-(n-2)(n-1)n\rho_n)=0.
\end{split}
\label{6}
\end{equation}

Let us introduce the generating function for these coefficients according
the definition: $G(u)=\sum_{n=0}^{\infty }\rho _{n}u^{n}$. Substituting this
expression into the Eq.~(\ref{6}) we obtain the following third order differential
equation for the $G(u)$:

\begin{equation}
(1-u^3)\frac{d^3G}{du^3}+\varepsilon_2(u^2-1)\frac{d^2(u^2G)}{du^2}+2\varepsilon_1(1-u)\frac{dG}{du}%
=0.
\label{7}
\end{equation}

\renewcommand{\figurename}{FIG.}
\begin{figure}[!t]
\includegraphics[scale=0.525]{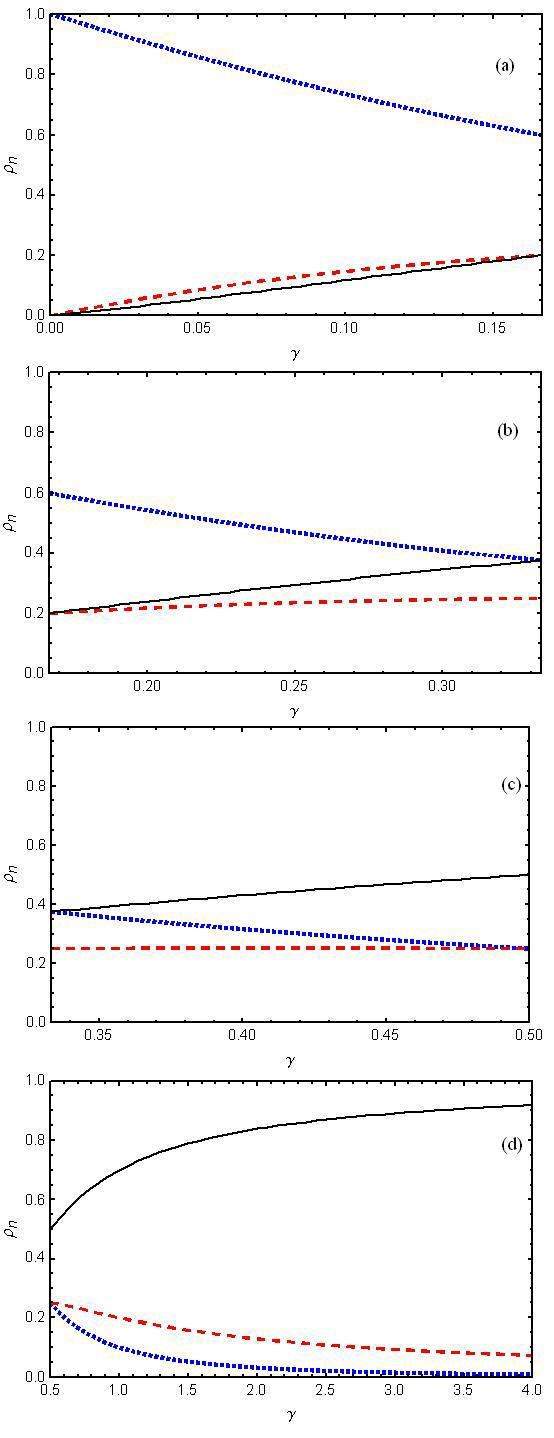}
\caption{(Color online) Population levels $\rho _{0}$ (spotted blue curves), $\rho _{1}$ (dashed red curves) and $\rho _{2}$ (black solid curves) as the function of single physical constant $\gamma $. These four regimes correspond to
the four ranges of $\gamma$: (a) $\gamma \le\frac{1}{6}$, then $\rho _{0}\ge\rho
_{1}\ge\rho _{2}$; (b) $\frac{1}{6}\le\gamma \le\frac{1}{3}$, then $\rho _{0}\ge\rho
_{2}\ge\rho _{1}$; (c) $\frac{1}{3}\le\gamma \le\frac{1}{2}$, then $\rho _{2}\ge\rho
_{0}\ge\rho _{1}$; (d) $\gamma \ge\frac{1}{2}$, then $\rho _{2}\ge\rho _{1}\ge\rho _{0}$.}
\end{figure}

It is impossible to find out analytical solution of Eq.~(\ref{7}) in
analytical form therefore we restrict ourselves to the case when
coefficients $\varepsilon _{1}$ and $\varepsilon _{2}$ are small but their
ratio can be of arbitrary value namely $\frac{\varepsilon _{2}}{%
2\varepsilon _{1}}=\gamma $. In the lowest approximation (when both $%
\varepsilon _{1}$ and $\varepsilon _{2}$ tend to zero), $G(u)$ is a certain
polynomial of the second order: $G_{0}(u)=\rho _{0}+\rho _{1}u+\rho _{2}u^{2}$%
, where populations $\rho _{n}$ should be found as follows. Substituting the
expression for $G(u)$ in Eq.~(\ref{7}) and taking into account that all $\rho _{i}=0
$ when $i>2$, and by virtue of normalization condition $\rho _{0}+\rho
_{1}+\rho _{2}=1$ we obtain the closed system of equations for the nonzero
coefficients $\rho _{n}$ that takes the form:

\begin{equation}
\left\{ 
\begin{array}{rcl}
\rho _{1}=2\gamma \rho _{0},   &  & \\ 
\rho _{2}=(6\gamma +1)\frac{\rho _{1}}{2}, &  &  \\ 
\rho _{0}+\rho _{1}+\rho _{2}=1. &  &  \\ 
&  & 
\end{array}%
\right.
\label{8}
\end{equation}

The solution of Eq.~(\ref{8}) looks as follows:

\begin{equation}
\left\{%
\begin{array}{rcl}
\rho_0=\frac{1}{6\gamma^2+3\gamma+1}, &  &  \\ 
\rho_1=\frac{2\gamma}{6\gamma^2+3\gamma+1}, &  &  \\ 
\rho_2=\frac{6\gamma^2+\gamma}{6\gamma^2+3\gamma+1}. &  &  \\ 
&  & 
\end{array}%
\right.
\label{9}
\end{equation}

Having in hands this solution we can analyze possible regimes of behavior
for AO in hard excitation regime as the function of the parameter $\gamma $. First
of all let us clarify two limiting cases a) $\gamma \rightarrow 0$ and b) $\gamma \rightarrow \infty$.

In the case a) $\rho _{0}\longrightarrow 1,\rho _{1}$and $\rho _{2}$ tend to
zero.This case corresponds to the vacuum state of AO in hard excitation regime
(or the state of rest in the classical case).

In the case b) $\rho _{0}=\rho _{1}=0,$and $\rho _{2}=1$. It is the case of
maximum possible excitation of the system in our approximation. It
corresponds to the state above threshold in classical case.

Now one can specify the four distinct regimes of the AO under study
depended on the parameter $\gamma=\frac{\varepsilon _{2}}{2\varepsilon _{1}}$. These regimes are represented in Fig.~1. 

It is interesting to compare the results obtained in the present paper with
similar ones relating to AO in soft excitation regime. Remind that
generation function G(u) for stationary states of AO in soft excitation regime
satisfies to the following second order differential equation (see Eq.~(26) in~Ref.~\onlinecite{1}):

\begin{equation}
(1+u)\frac{d^{2}G}{du^{2}}-\nu u\frac{dG}{dt}-\nu G(u)=0,
\label{10}
\end{equation}

where $\nu$  is the only nonlinear parameter of this oscillator. Its solution
that satisfies all physical conditions can be expressed as

\begin{equation}
G(u)=\frac{F(1,\nu ,\nu (1+u))}{F(1,\nu ,2\nu )},
\label{11}
\end{equation}

where $F(a,b,x)$ is the standard confluent hyper-geometric function. Using
the expansion of this function namely: $F(a,b,x)=1+(\frac{a}{b})x+\frac{%
a(a+1)x}{2!b(b+1)}+...$ one can easily see that if parameter of nonlinearity 
$\nu $ tends to zero corresponding generation function tends to:

\begin{equation}
G_0(u) \simeq\frac{2+u}{3}
\label{12}
\end{equation}

Thus the AO in soft excitation regime and small nonlinearity reduced to the
two level system with population $\frac{2}{3}$ in the lower and $\frac{1}{3}$
in the upper level respectively. We see that compared with such primitive
regime the case AO in hard excitation regime reveals considerably much more rich
behavior.

Let us sum up: the quantum model of an AO in hard excitation regime is
firstly proposed in this paper. Using the methods of the quantum theory of
the OQS, the Lindblad equation for the density matrix of the oscillator was
obtained, and it was used to find a solution for the populations of the
stationary states of the oscillator in the case when the physical parameters
of the model are small. It was shown that the quantum model proposed here
has much more rich behavior then AO in soft excitation regime. In conclusion
it is worth to note that the model AO in hard excitation regime considered in
present paper, if it should be implemented as physical device, naturally
realizes the curious case of three level quantum system in which one can
achieves (by varying only single parameter) population inversion on any
desired pair of levels.

\renewcommand\refname{}

\righthyphenmin=3

\end{document}